# Investigations into X-band dielectric disk accelerating structures for future linear accelerators

Yelong Wei [1,*] and Alexej Grudiev [1]

[1]*CERN, Geneva CH-1211, Switzerland*

Abstract: Dielectric disk accelerating (DDA) structures are being studied as an alternative to conventional disk-loaded copper structures. This paper investigates numerically an efficient X-band DDA structure operating at a higher order mode of $TM_{02}$-π. This accelerating structure consists of dielectric disks with irises arranged periodically in a metallic enclosure. Through optimizations, the RF power loss on the metallic wall can be significantly reduced, resulting in an extremely high quality factor $Q_0 = 111064$ and a very high shunt impedance $R_{\text{shunt}} = 605$ MΩ/m. The RF-to-beam power efficiency reaches 46.6% which is significantly higher than previously-reported CLIC-G structures with an efficiency of only 28.5%. The optimum geometry of the regular and the end cells is described in detail. Due to the wide bandwidth from the dispersion relation of the accelerating mode, the DDA structure is allowed to have a maximum number of 73 regular cells with a frequency separation of 1.0 MHz, which is superior to that of conventional RF accelerating structures. In addition, the DDA structure is found to have a short-range transverse wakefield lower than that of the CLIC-G structure.

## I. Introduction

Over the past few decades, conventional disk-loaded copper radiofrequency (RF) structures have been widely studied, and used to accelerate particles in a variety of applications for scientific research [1-2], medical cancer therapy [3-4], and industrial processing [5-6]. A high accelerating gradient of up to 100 MV/m has been demonstrated at room temperature for an X-band copper structure, which has been studied as the baseline accelerating structure design for the Compact Linear Collider (CLIC) main linac [7-11]. Such a structure is named 'CLIC-G', operating at 11.994 GHz in 2π/3 mode with an unloaded quality factor $Q_0 \approx 5600$, a shunt impedance $R_{\text{shunt}} \approx 92$ MΩ/m, and a power efficiency $\eta_{\text{rf-beam}} \approx 28.5\%$. However, these normal-conducting RF structures are substantially less favourable in terms of RF-to-beam power efficiency when compared to superconducting RF cavities realizing a much higher quality factor and a higher shunt impedance, although accelerating gradient is higher in these normal-conducting structures. Thus the major challenge for a room-temperature RF accelerating structure is how to achieve a quality factor and a shunt impedance much higher than those of conventional normal-conducting RF structures. This is of particular importance for future linear accelerators.

A potential alternative to the conventional disk-loaded copper structure is a dielectric loaded accelerating (DLA) structure [12-16], which utilizes dielectrics to slow down the phase velocity of travelling wave in the beam channel. A DLA structure comprises a simple geometry where a uniform and linear dielectric tube surrounded by a copper cylinder. Simulation studies [17] have shown that the ratio of the peak electric field to the average accelerating field in a DLA structure is about one, indicating that the accelerating gradient achieved is potentially higher than that of conventional copper structures assuming dielectrics and metals have the same breakdown limit. The DLAs also have another advantage in terms of the ease of applying damping schemes for beam-induced deflection modes [18-19], which can cause bunch-to-bunch beam breakup and intrabunch head-tail instabilities [20].

The DLA structures were initially proposed in the 1940s [21-24], and experimentally demonstrated in the 1950s [25-27]. Since that time, disk-loaded metallic structures have prevailed for accelerator research and development because of their high quality factor and high field holding capability. Thanks to remarkable progress in new ceramic materials with high dielectric permittivity ($\varepsilon_r > 20$), low loss ($\tan\delta \leq 10^{-4}$) [28-30], and ultralow-loss ($\tan\delta \leq 10^{-5}$) [31-32], studies on DLA structures are gradually being revived. For example, fused silica, chemical vapor deposition (CVD) diamond, alumina and other ceramics have been proposed as materials for DLA structures [33-35], and experimentally tested with high-power wakefield accelerating structures at Argonne National Laboratory [36-39]. In the last two decades, different kinds of DLA structures with improved performance have been reported, such as a dual-layered dielectric structure [40], a hybrid dielectric and iris-loaded accelerating structure [41], a multilayered dielectric structure [42], a disk-and-ring tapered accelerating structure [43], and a dielectric disk accelerating (DDA) structure [44]. Among these DLA structures, a dielectric assist accelerating structure proposed by Satoh *et al.* [45-46] is of particular interest because it realized an extremely high quality factor and a very high shunt impedance at a C-band frequency. Building on these developments, a DDA structure operating at a high frequency (X-band) appears to be very promising for future linear accelerators considering a high accelerating gradient of up to 100 MV/m demonstrated at an X-band frequency [8-9].

*yelong.wei@cern.ch

In this paper we explore numerically an efficient X-band DDA structure operating at $TM_{02}$-π mode. The RF power loss on the conducting wall can be greatly reduced by adjusting the electromagnetic field distribution for the accelerating mode of $TM_{02}$-π, thereby realizing an extremely high quality factor and a very high shunt impedance at room temperature. Thus, the RF-to-beam power efficiency can also be dramatically improved. Section II presents detailed optimization studies for a regular cell and an end cell in the DDA structures. Section III describes the accelerating performance for a DDA structure with a number of regular cells and two end cells. Section IV shows the simulated short-range and long-range transverse wakefields for the DDA structure, and a comparison with the CLIC-G structure is also made.

# II. Geometry optimization

The DDA structures have the same geometry as the dielectric assist accelerating structures proposed by Satoh *et al.* [45-46]. They consist of dielectric cylinders with irises, periodically arranged in a metallic enclosure, and operate at a $TM_{02}$ mode with a $\pi$ phase advance (standing wave). The ceramic materials for the dielectric-based accelerating structures have to withstand high accelerating fields, prevent potential charging by particle beams, have good thermal conductivity, and generate low power loss. $TiO_2$-doped $Al_2O_3$ ceramic, with a relative permittivity $\varepsilon_r = 9.64$ and an ultralow loss tangent $\tan\delta = 6 \times 10^{-6}$, which has been demonstrated by [46], is chosen as the dielectric material for our DDA structures. As illustrated in Fig. 1, the geometry of a DDA structure is comprised of two types of cell structure: a regular cell and an end cell, which are represented by red and blue areas, respectively. The regular cell provides the particle beam with the accelerating field, while the end cell serves to reduce the RF power dissipation on the surface of both the conducting end plates. The following geometry optimization aims to maximize the unloaded quality factor $Q_0$ and the shunt impedance $R_{\text{shunt}}$ and for both regular and end cells.

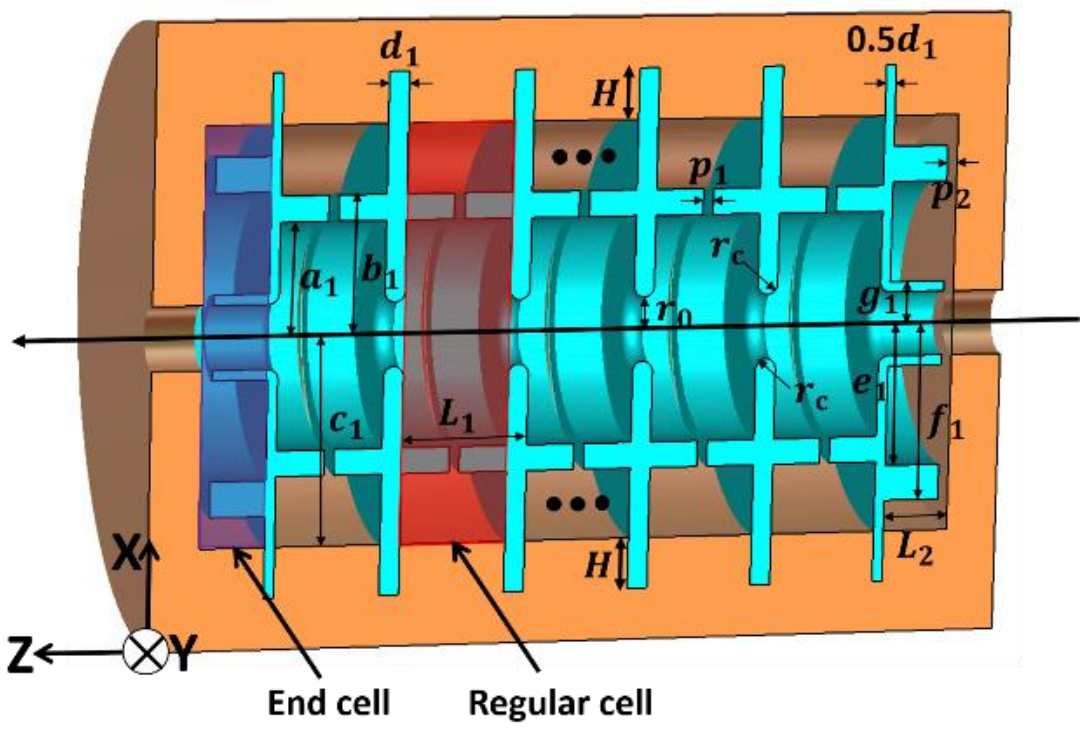

Figure 1. Conceptual schematic of an X-band DDA structure. $L_1$, $H$, $r_0$, $r_c$, $a_1$, $b_1$, $c_1$, $d_1$, $p_1$ represent the periodic length, dielectric insertion depth, iris radius, corner fillet radius, inner radius, outer radius, copper waveguide radius, dielectric disk thickness, and vacuum gap, respectively, for the regular cells while $L_2$, $e_1$, $f_1$, $g_1$, $p_2$ represent the length, inner radius, outer radius, iris thickness, and vacuum gap, respectively, for the end cells.

## A. Regular cell

The accelerating fields are distributed in the regular cell, which dominates the RF parameters such as the unloaded quality factor and the shunt impedance for a DDA structure. Figure 2 shows the axially symmetric cross-section geometry for a regular cell, which consists of the iris radius $r_0$, the corner fillet radius $r_c$, the inner radius $a_1$, the outer radius $b_1$, the copper waveguide radius $c_1$, the dielectric disk thickness $d_1$, the vacuum gap $p_1$, and the dielectric insertion depth $H$ at a constant periodic length $L_1$. After selecting $L_1, r_0, r_c, p_1, H$, the combination of $a_1$, $b_1$, $c_1$, and $d_1$ primarily determines the dispersion relation of the accelerating mode of $TM_{02}$-π. HFSS [47] is used to compute the electromagnetic fields in this regular cell structure. In this subsection optimization studies are carried out to maximize the unloaded quality factor and the shunt impedance for a regular cell operating at an X-band frequency $f_0 = 11.994$ GHz.

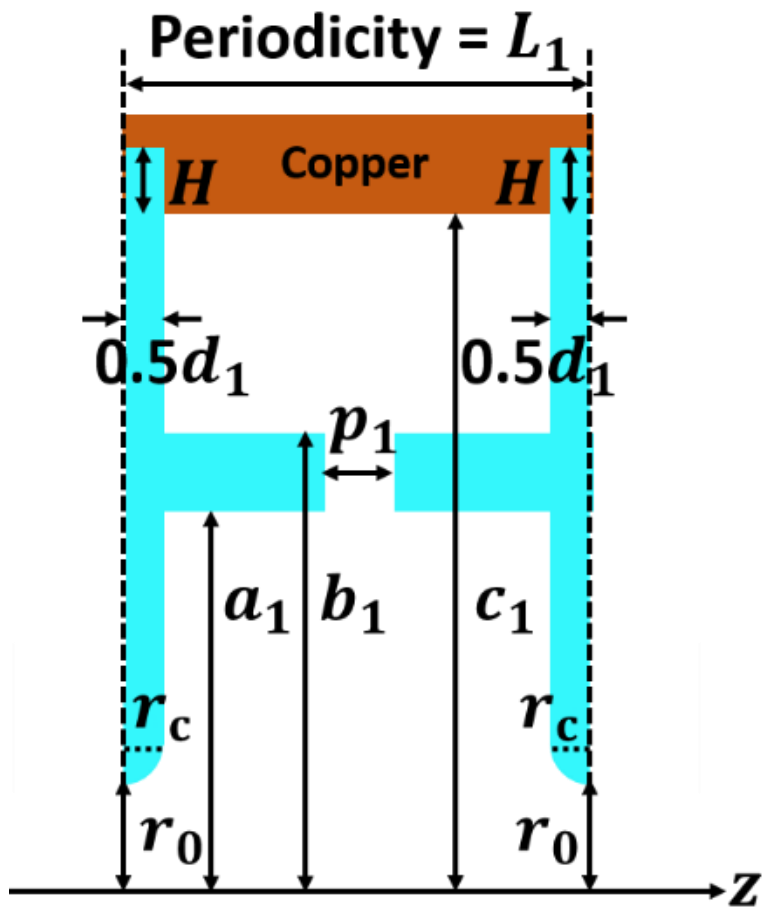

Figure 2. Longitudinal cross-section geometry of a regular cell.

Consideration of such a regular cell is used to generate a $TM_{02}$-π mode, in which the periodic length $L_1$ is equal to $L_1 = \lambda_0/2$, where $\lambda_0$ is the free-space wavelength for $f_0$. The iris aperture radius $r_0$ is chosen as $r_0 = 3.15$ mm which is comparable with that of CLIC-G structure [7-11] considering the DDA structure can be potentially used for CLIC main linac in the future. According to the optimization processing in Ref [45], the disk thickness $d_1$ is chosen to be a quarter of the resonant wavelength in such a dielectric material. We can therefore get the optimum thickness $d_1 = \lambda_0/(4\sqrt{\epsilon_r}) \approx 2.0$ mm. A fillet radius $r_c = d_1/2 = 1.0$ mm for the dielectric corner is added to reduce $E_p/E_{acc}$, where $E_p$ and $E_{acc}$ are the peak electric fields and the average accelerating gradient in a regular cell. A vacuum gap $p_1 = d_1/4 = 0.5$ mm is added between each regular cell for evacuating the whole vacuum region for the regular cells from the beam hole. An additional dielectric layer is inserted into the outer copper wall with a depth of $H = 2.5\, d_1 = 5.0$ mm, which improves the thermal contact. After selection of $L_1, r_0, d_1, r_c, p_1, H$, the resonant frequency $f_0$ is determined primarily by the combination of $a_1$, $b_1$, and $c_1$. When $a_1$ and $b_1$ are fixed, the value of $b_1$ can be calculated for the given frequency $f_0$. Based on this reasoning, the unloaded quality factor and the shunt impedance for the regular cell are calculated by sweeping through different values of $a_1$ and $c_1$. It should be noted here that in these simulations the periodic boundary conditions are applied to such a regular cell.

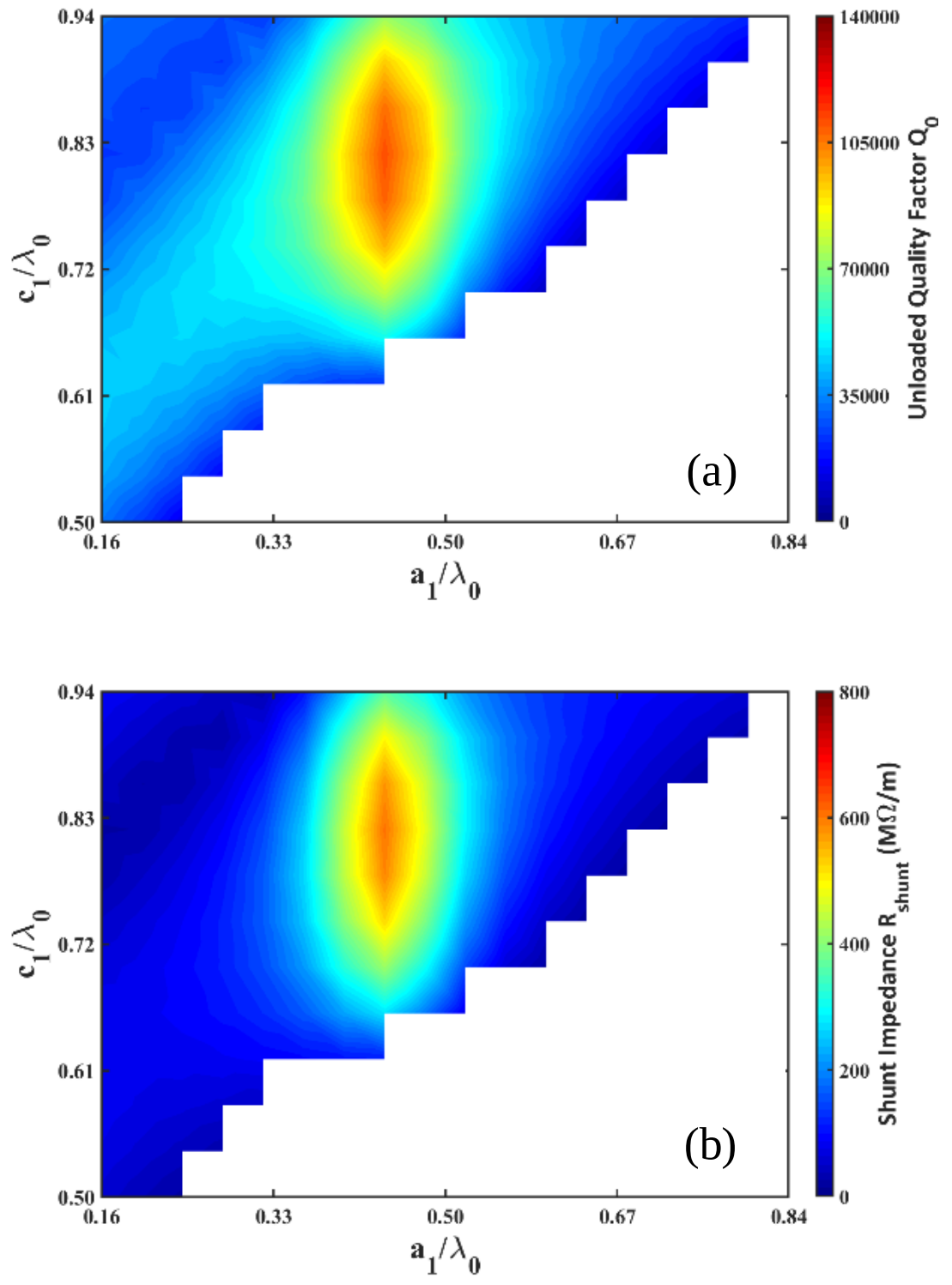

Figure 3. The calculated unloaded quality factor $Q_0$ (a), and shunt impedance $R_{shunt}$ (b) as a function of geometrical parameters of $a_1$ and $c_1$ for a regular cell

Table 1. Optimum parameters for a regular cell of DDA operating at $TM_{02}$-π mode

| Geometry | Regular cell |
|---|---|
| Dielectric constant $\varepsilon_r$ | 9.64 |
| Dielectric loss tangent tan $\delta$ | $6\times10^{-6}$ |
| Iris radius $r_0$ [mm] | 3.15 |
| Corner fillet radius $r_c$ [mm] | 1.0 |
| Inner radius $a_1$ [mm] | 11.1 |
| Outer radius $b_1$ [mm] | 13.626 |
| Copper waveguide radius $c_1$ [mm] | 20.5 |
| Dielectric disk thickness $d_1$ [mm] | 2.0 |
| Vacuum gap $p_1$ [mm] | 0.5 |
| Dielectric insertion depth $H$ [mm] | 5.0 |
| Periodical length $L_1$ [mm] | 12.5 |
| Phase advance | $180^0$ |
| Acceleration mode | $TM_{02}$ |
| Frequency $f$ [GHz] | 11.994 |
| Unloaded $Q_0$ | 111064 |
| Shunt impedance $r'$[MΩ/m] | 605 |

Figure 3 shows the dependence of the unloaded quality factor $Q_0$ and the shunt impedance $R_{\text{shunt}}$ on the geometrical parameters $a_1$ and $c_1$, respectively, for a regular cell. The white areas shown in Fig. 3 indicate the parameter combinations where an accelerating mode of $TM_{02}$-$\pi$ cannot be found at the operating frequency. As illustrated in Fig. 3, a maximum unloaded quality factor $Q_0 = 111056$ and a maximum shunt impedance $R_{\text{shunt}} = 605$ MΩ/m can be realized for a regular cell with geometrical parameters $a_1 = 0.444\lambda_0$ and $c_1 = 0.82\lambda_0$. The optimum parameters are listed in Table 1. It is found that such an unloaded quality factor and a shunt impedance are 15 and 5 times higher, respectively, compared to those of CLIC-G structures. Considering that these optimum geometry parameters are thick enough for fabrication with good precision, the regular cell can realistically be built.

The ratio of the peak electric field $E_{\text{p}}$ to the average accelerating field $E_{\text{a}}$ limits the achievable accelerating gradient for conventional iris-loaded metallic structures. Typically this ratio $E_{\text{p}}/E_{\text{a}} \geq 2$ [7, 10-11, 48-49]. Figure 4 shows the electric field distribution $E/E_{\text{a}}$ and magnetic field distribution $H/E_{\text{a}}$ of the $TM_{02}$-$\pi$ mode in a regular cell of DDA structures, where $E$, $E_{\text{a}}$, $H$ represent the electric field, the average accelerating field, and the magnetic field, respectively. These simulation results indicate that the ratio of the peak electric field to the average accelerating field is 2.07, while the ratio of the peak magnetic field to the average accelerating field is 3.49 mA/V for the regular cell, which are comparable to those of CLIC-G structures at the same operating frequency. It is also shown in Fig. 4 that most of the RF power is stored in the vacuum region. In this case, the power loss on the conducting cylinder surface can be dramatically reduced, resulting in an extremely high quality factor and a very high shunt impedance for a regular cell at room temperature.

In summary, after optimization, the regular cell of DDA structures has superior RF parameters in terms of an extremely high unloaded quality factor and a very high shunt impedance, while keeping $E_{\text{p}}/E_{\text{a}}$ and $H_{\text{p}}/E_{\text{a}}$ similar to the existing CLIC-G structure.

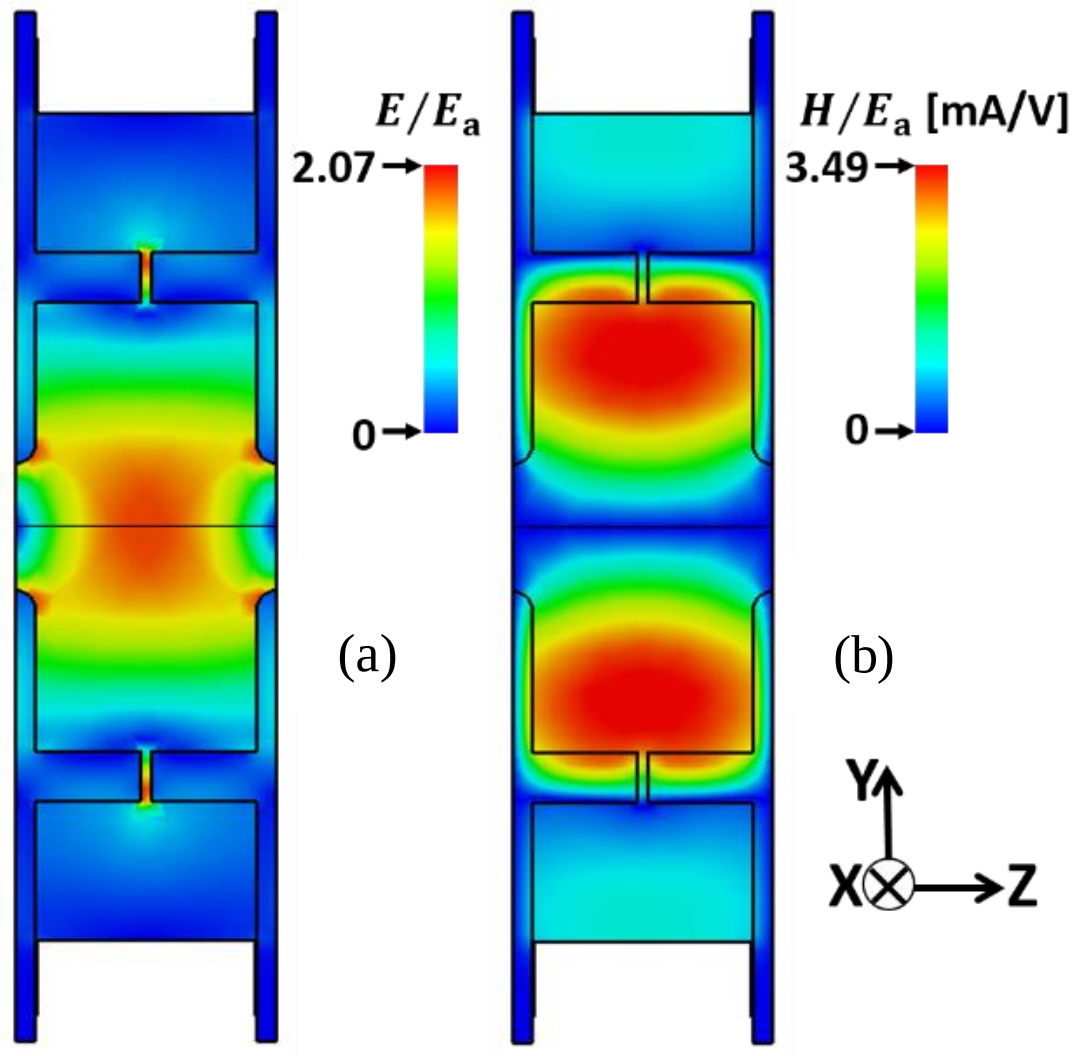

Figure 4. (a) Electric field distribution $E/E_{\text{a}}$, and (b) magnetic field distribution $H/E_{\text{a}}$ for the accelerating mode of $TM_{02}$-$\pi$ in the regular cell.

## B. End Cell

In order to reduce the RF power loss dissipated on the surfaces of both conducting end plates, two end cells are added into the DDA structures. Figure 5 shows the longitudinal cross-section geometry of a single-cell DDA structure with a regular cell and two end cells. There are five geometrical parameters $L_2$, $e_1$, $f_1$, $g_1$, $p_2$ representing the length, the inner radius, the outer radius, the iris thickness, and the vacuum gap, respectively, for an end cell. In order to evacuate the whole vacuum region in the regular and end cells from the beam hole, a vacuum gap $p_2 = p_1/2$ is chosen for the end cells. After fixing $p_2$, the best RF parameters including the unloaded quality factor and the shunt impedance can be obtained by tuning the combination of $L_2$, $e_1$, $f_1$, and $g_1$ for the desired frequency $f_0 = 11.994$ GHz.

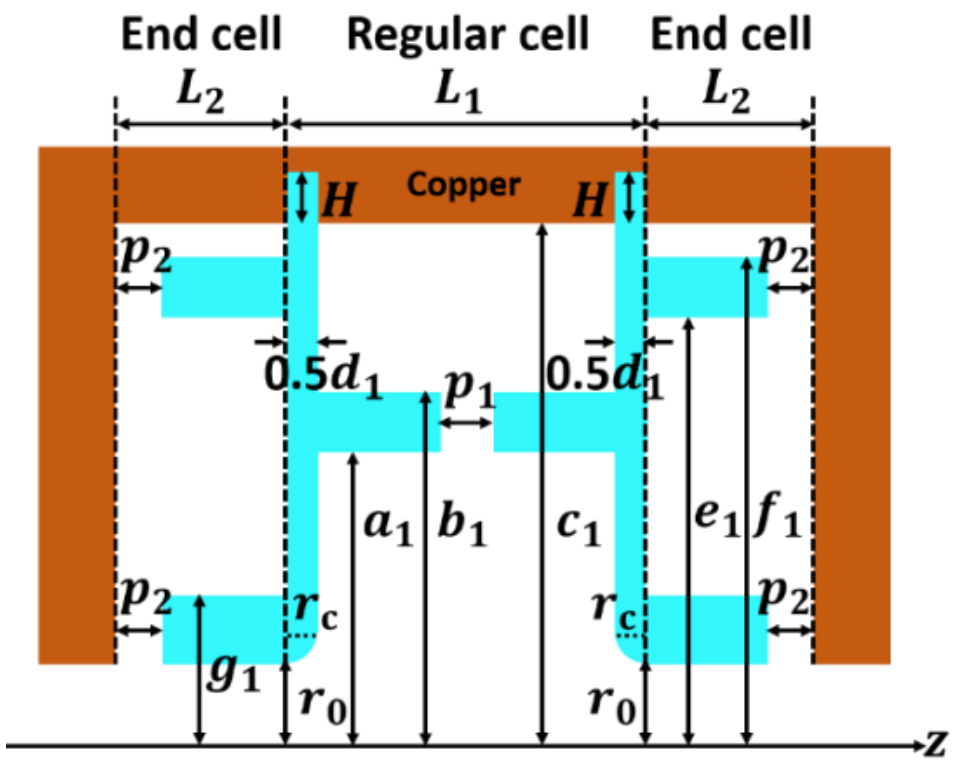

Figure 5. Longitudinal cross-section geometry of a single-cell DDA structure with a regular cell and two end cells.

For a regular cell which is electrically shorted just by the conducting end plates, without dielectric end-cells, simulations show that the RF power loss on the dielectric accounts for 2.9% of total power loss while the remaining 97.1% of power loss comes from the conducting cylinder surface. The ratio of RF power loss on the conducting cylinder surface to the total power loss can be reduced to 89.6% by adding two dielectric end cells into a DDA structure, as shown in Fig. 5. This means that the total power loss in a DDA structure with end cells can be about 70% lower than that of a DDA structure without end cells. The magnetic field distribution determines the RF power loss for a single-cell DDA structure with a regular cell and two end cells, and is plotted in Fig. 6. This shows that most of the RF power is still concentrated in the vacuum region of regular cells, so the power loss on the conducting end plates is greatly reduced by using the end cells.

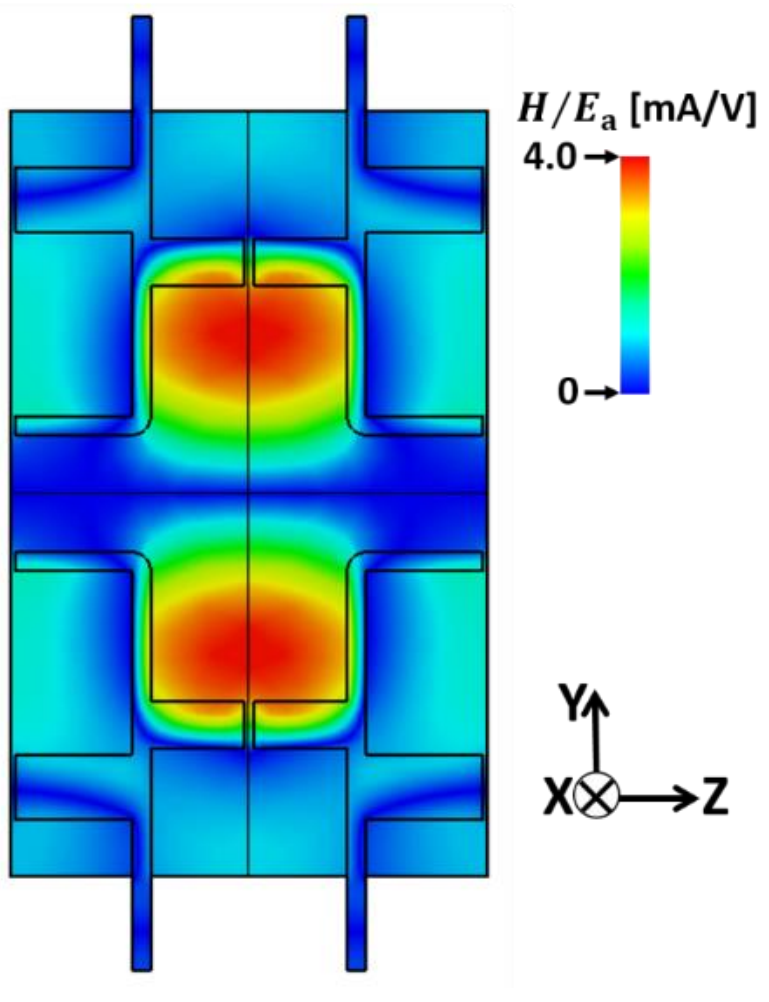

Figure 6. Magnetic field distribution for a single-cell DDA structure with a regular cell and two end cells.

A cylinder dielectric with a thickness of 1 mm ($g_1$ = 4.15 mm) is chosen for the optimizations which follow. Thus the combination of geometrical parameters $L_2$, $e_1$, and $f_1$ determines the frequency and RF performance of the accelerating mode which is $TM_{02}$-$\pi$. Using the same optimization procedure as regular cells, we sweep through different $L_2$ and $e_1$values to get the desired frequency $f_0 = 11.994$ GHz. Figure 7 shows the dependence of the unloaded quality factor $Q_0$ and the shunt impedance $R_{shunt}$ on the geometrical parameters $L_2$ and $e_1$, respectively. As Fig. 7 shows, a maximum unloaded quality factor of $Q_0 = 52174$ and a maximum shunt impedance of $R_{shunt} = 169$ MΩ/m can be realized for a single-cell DDA structure with $e_1 = 0.56\lambda_0$, $L_2 = 0.26\lambda_0$, and $f_1 = 0.70\lambda_0$.

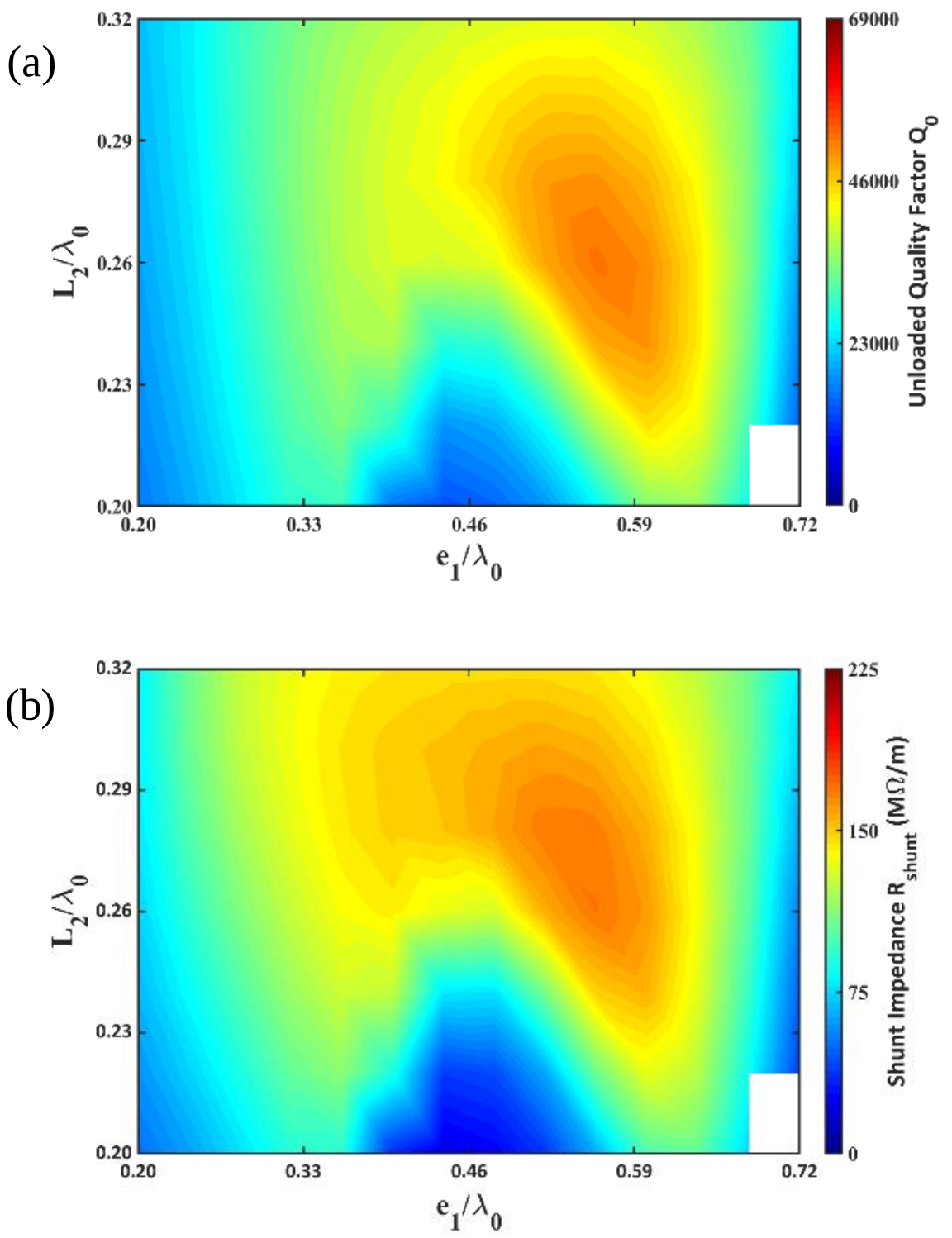

Figure 7. The calculated unloaded quality factor $Q_0$ (a), and shunt impedance $R_{shunt}$ (b) as a function of geometrical parameters $e_1$ and $L_2$ for a single-cell DDA structure.

# III. Full structure performance

After the optimization studies, we move on to study the properties of a whole DDA structure with a number of regular cells and two end cells. This section investigates the accelerating performance of a DDA structure, including the dispersion relation of the accelerating mode of $TM_{02}$-$\pi$, the dependence on the dielectric loss tangent, and the RF-to-beam power efficiency.

## A. Dispersion relation

In a RF accelerating structure the precise relationship between angular frequency $\omega$ and wave number $k_z$ is called the dispersion relation. Figure 8 shows the dispersion relation of the accelerating mode $TM_{02}$ in a regular cell. Such a regular cell has the geometry parameters listed in Table 1. As shown in Fig. 8, the red curve can be approximated by

$$f = f_{\mathrm{int}}/\sqrt{1 + k\cos\Delta\varphi}, \quad (1)$$

where $f_{\mathrm{int}} = \sqrt{2}f_\pi f_0/\sqrt{f_\pi^2 + f_0^2}$ is called the intrinsic frequency, $k = (f_\pi^2 - f_0^2)/(f_\pi^2 + f_0^2)$ is called the coupling coefficient, $\Delta\varphi$ is the phase advance, and $f_\pi$ and $f_0$ are corresponding frequencies for a phase advance of $\pi$ and 0, respectively. By substituting $f_\pi = 11.994$ GHz and $f_0 = 9.755$ GHz into these equations, we get an intrinsic frequency $f_{\mathrm{int}} = 10.703$ GHz and a coupling coefficient $k = 0.204$ . A bandwidth $BW = f_{\mathrm{int}}k = 2.18$ GHz is then obtained. For a standing-wave accelerating structure the overlap between adjacent modes is an issue from the tunability and operational point of view. To avoid mode overlapping [50-52] in a periodical RF accelerating structure (which here denotes the DDA structure), the number of regular cells $N < \sqrt{Q_\pi \pi^2 k/4} = 236$ , where $Q_\pi$ is the unloaded quality factor and is calculated to be 111064 for operating $\pi$ mode. In this case, the maximum number of regular cells is $N$ =235. However, the corresponding frequency separation for the operating $\pi$ mode between the $N-1$ cell and the $N$ cell is calculated to be $\Delta f_{\mathrm{N-1,N}}^{\pi} = f_{\mathrm{int}}k(\pi/2N)^2 =$ 0.1 MHz which is very sensitive to tuning system. In order to have a frequency separation of 1 MHz with a moderate sensitivity to tuning system, the maximum number of regular cells is calculated to be $N$ =73. In summary, the DDA structure operating at $TM_{02}$-$\pi$ mode is analytically allowed to have a maximum number of 73 regular cells with a frequency separation of 1.0 MHz.

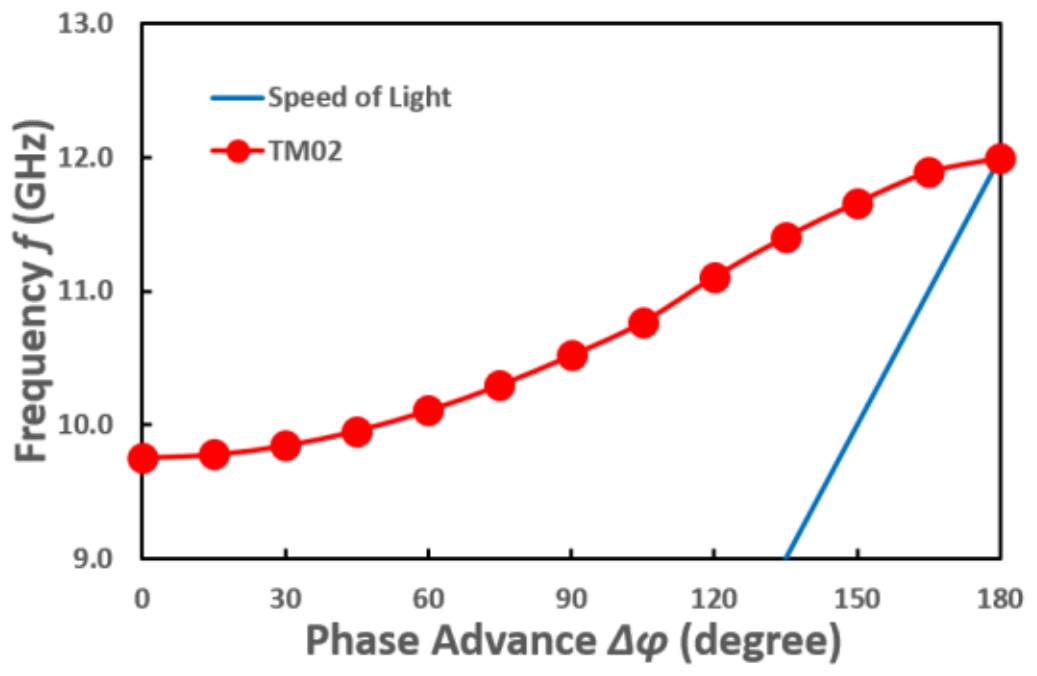

Figure 8. Dispersion curves of the accelerating mode $TM_{02}$ (red curve) and speed of light (blue curve).

With an increasing number of regular cells, more RF power is concentrated in the regular cells while less RF power is stored in the end cells. In this case the RF power loss in the end cell gradually decreases, thereby increasing the unloaded quality factor and shunt impedance for a whole DDA structure. As shown in Fig. 9, the simulated unloaded quality factor and shunt impedance of a DDA structure gradually increase and become saturated to those of a regular cell with periodic boundary conditions. When the number of regular cells for a DDA structure is more than 13, the unloaded quality factor and shunt impedance become constant, at $Q_0 =$ 111064 and $R_{\mathrm{shunt}} = 605$ MΩ/m, respectively. In other words, the end cells have negligible effect on the RF performance for a DDA structure with a larger number of regular cells. It should be noted here that the end cells with different geometries are applied for the simulation results as seen in Fig. 9.

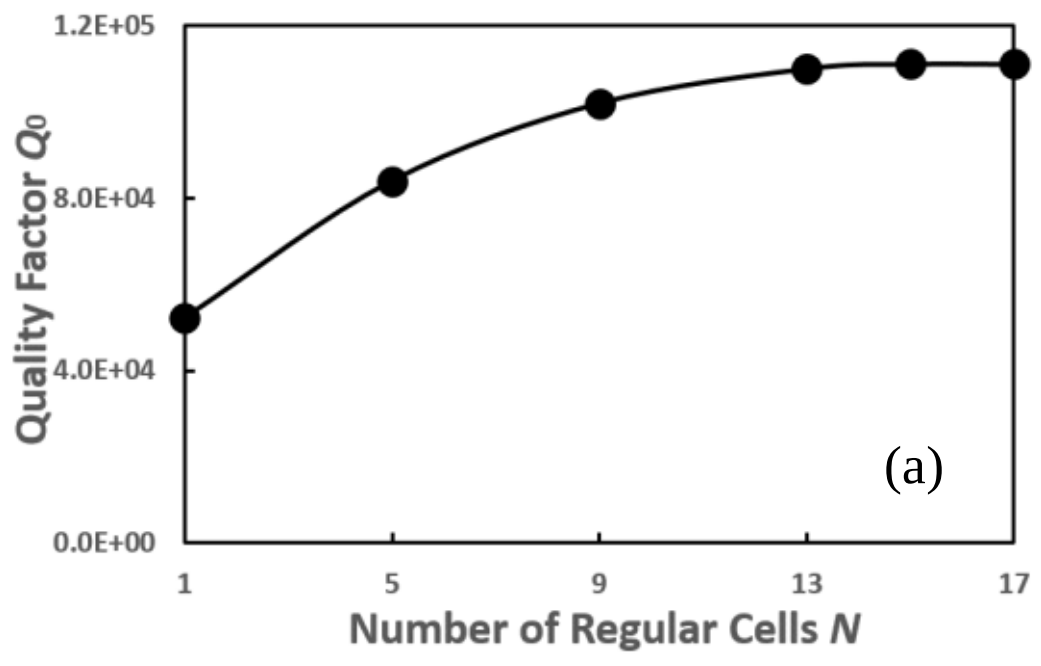

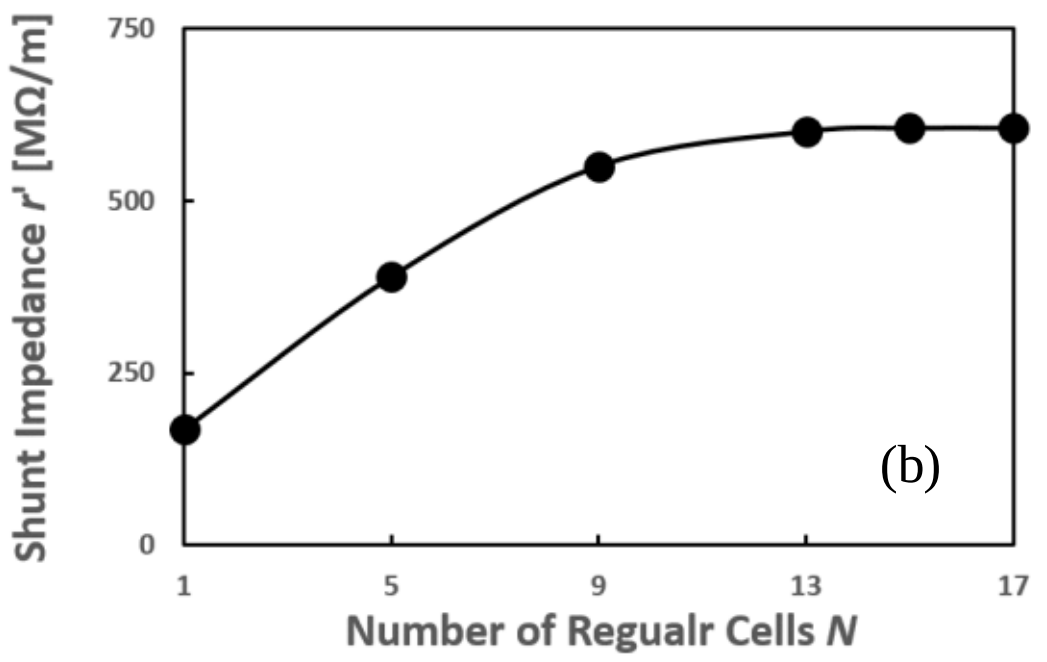

Figure 9. Dependence of the unloaded quality factor $Q_0$ (a), and shunt impedance $R_{\mathrm{shunt}}$ (b) on the number of regular cells for a DDA structure.

## B. Dielectric loss tangent

Different ceramic materials with a low loss tangent ($\tan\delta \leq 10^{-4}$) [28-30], and an ultralow loss tangent ( $\tan\delta \leq 10^{-5}$ ) [31-32], have been intensively studied for DLAs over the past few decades. As discussed in Section II, the RF parameters listed in Table 1 are calculated using the dielectric constant $\varepsilon_r = 9.64$ and the loss tangent $\tan\delta = 6 \times 10^{-6}$ of $TiO_2$-doped $Al_2O_3$ ceramics [46]. In realistic situations the loss tangent may vary and depends strongly on the manufacturing process for the ceramic materials. Thus it is of great importance to study how the RF parameters (the unloaded quality factor and shunt impedance) vary with different loss tangents, for a DDA structure. The unloaded quality factor and shunt impedance vary as the number of regular cells increases and remain unchanged when the number of regular cells is larger than 13, as shown in Fig. 9. We therefore only need to simulate the dependence of the unloaded quality factor and shunt impedance on the dielectric loss tangent for a regular cell which has similar RF parameters to a DDA having more than 13 regular cells.

Figure 10 shows the relationship between the unloaded quality factor and shunt impedance and the dielectric loss tangent for a DDA structure. As Fig. 10 shows, both the unloaded quality factor and the shunt impedance increase when the loss tangent is decreased. These RF parameters gradually become saturated when the loss tangent is smaller than $10^{-6}$. In this situation the RF power loss in the dielectric material is negligible compared to that on the conducting wall. A maximum unloaded quality factor $Q_0 = 142611$ and shunt impedance $R_{\mathrm{shunt}} = 776.5$ MΩ/m can be achieved when the loss tangent is below $10^{-7}$. $TiO_2$-doped $Al_2O_3$ ceramic with a loss tangent of $10^{-5}$ can also be obtained in a relatively simpler manufacturing process compared to achieving an ultralow-loss tangent of $6 \times 10^{-6}$, which requires complicated treatment. The corresponding unloaded quality factor and shunt impedance are $Q_0 = 96767$, $R_{\mathrm{shunt}} = 527$ MΩ/m, respectively, for a loss tangent of $10^{-5}$, as shown in Fig. 10.

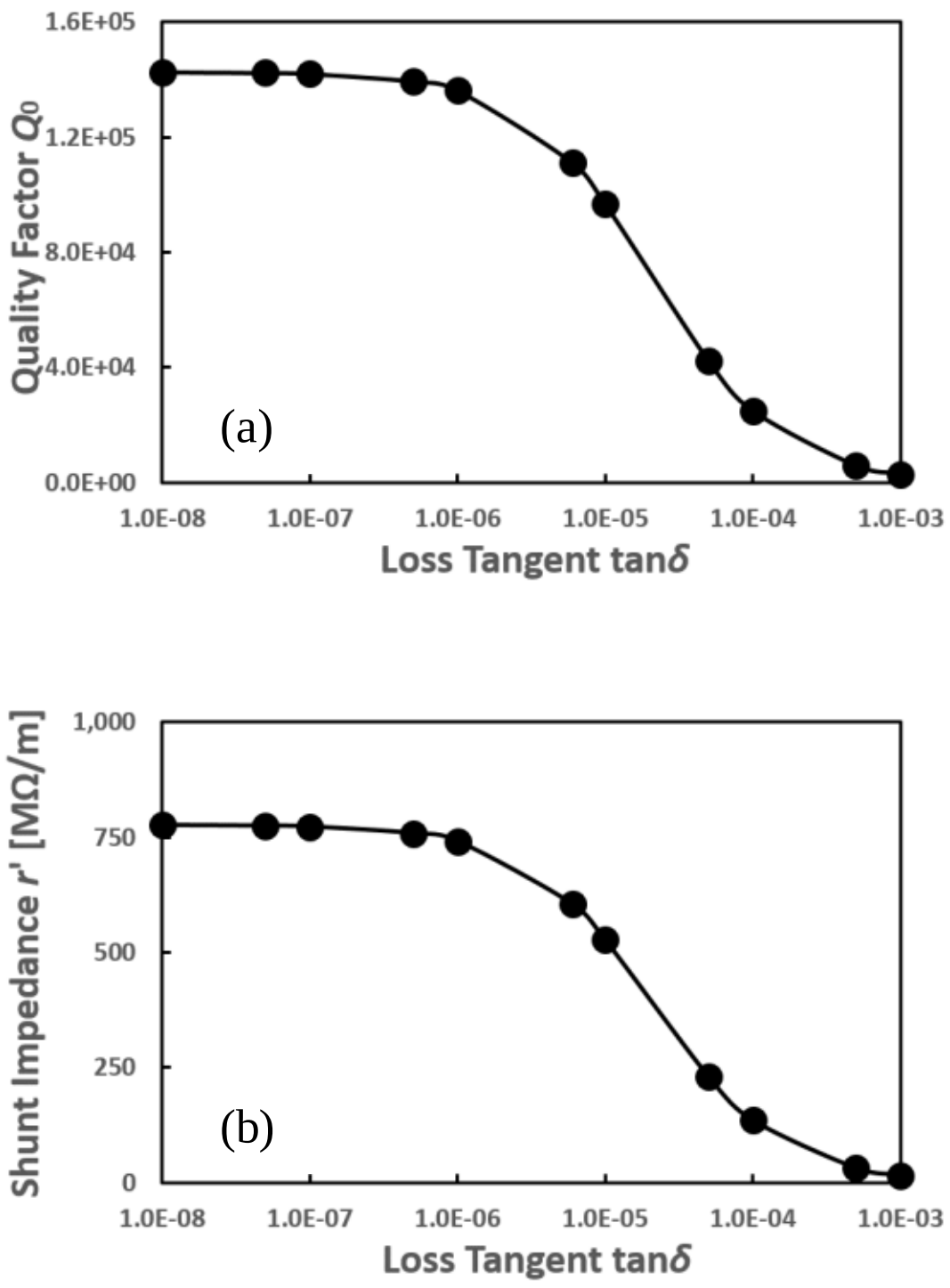

Figure 10. Dependence of the unloaded quality factor $Q_0$ (a), and shunt impedance $R_{\mathrm{shunt}}$ (b) on the dielectric loss tangent $\tan\delta$ for a DDA structure.

## C. RF-to-beam power efficiency

Of particular concern for a linear accelerator is its RF-to-beam power efficiency. The RF-to-beam power efficiency for a standing-wave accelerating structure (here it denotes a DDA structure) is defined by

$$\eta_{\mathrm{rf-to-beam}} = \left(\frac{P_{\mathrm{b}}}{P_{\mathrm{rf}}}\right) \times \left(\frac{T_{\mathrm{b}}}{T_{\mathrm{rf}}}\right), \tag{2}$$

where $P_{\mathrm{b}} = I_{\mathrm{b}} V_{\mathrm{a}}$ is the beam power, $I_{\mathrm{b}}$ is the beam current, $V_{\mathrm{a}}$ is the average accelerating voltage, $P_{\mathrm{rf}}$ is the input RF power, $T_{\mathrm{b}}$ is the period of time that beam appears, and $T_{\rm rf}$ is the total length of RF pulse. The RF power loss is given by

$$P_{\rm loss} = \frac{V_{\rm a}^2}{r'NL}, \quad (3)$$

where $V_{\rm a} = NLE_{\rm load}$ is the average accelerating voltage, $N$ is the number of cells, and $L$ is the length of a single cell, $E_{\rm load}$ is the average loaded accelerating gradient, $r'$ is the shunt impedance of an accelerating structure. The input RF power is

$$P_{\rm rf} = P_{\rm b} + P_{\rm loss} = I_{\rm b}V_{\rm a} + \frac{V_{\rm a}^2}{r'NL}. \quad (4)$$

So that we get

$$\frac{P_{\rm b}}{P_{\rm rf}} = \frac{I_{\rm b}}{E_{\rm load}/r'+I_{\rm b}}. \quad (5)$$

The unloaded gradient is defined by

$$E_{\rm unload} = \sqrt{P_{\rm rf}\, r'/(NL)} = \sqrt{I_{\rm b} r' E_{\rm load} + E_{\rm load}^2}. \quad (6)$$

Thus the loaded gradient is derived:

$$E_{\rm load} = E_{\rm unload}\left(1 - e^{-\frac{t}{\tau}}\right), \quad (7)$$

where $\tau = 2Q_0/[\omega(1+\beta)]$ is the time constant, $Q_0$ is the unloaded quality factor, $\omega$ is the angular frequency, $\beta = P_{\rm rf}/P_{\rm loss}$ is the coupling constant, $t$ is the time. The beam current $I_{\rm b} = 1.19$ A from the CLIC, and the loaded gradient $E_{\rm load} = 100$ MV/m, the unloaded quality factor $Q_0 = 111064$ and the shunt impedance $r' = 605$ MΩ/m from a DDA structure, are used for our calculations. By substituting these numbers into Eqs. (5), (6) and (7), we get $P_{\rm b}/P_{\rm rf} = 0.878$, $E_{\rm unload} = 286.35$ MV/m, $\tau = 320.4$ ns, and

$$100 = 286.35\left(1 - e^{-\frac{t}{\tau}}\right). \quad (8)$$

After solving Eq. (8), the calculated RF power filling time for a DDA structure is $T_{\rm fill} = 137.6$ ns, as shown in Fig. 11. The beam time is set to $T_{\rm b} = 155.6$ ns which is the same as that of the CLIC-G structure. By substituting these numbers into Eq. (2), a RF-to-beam power efficiency for a DDA structure as high as $\eta_{\rm rf-to-beam} = 46.6\%$ is reached, which is 63.5% higher than previously reported CLIC-G structures with an efficiency of 28.5%. It should be pointed out that this efficiency is based on a DDA structure with a ultralow-loss tangent of $6 \times 10^{-6}$. It decreases to 44.8% when the loss tangent is $1 \times 10^{-5}$. The comparison between the CLIC-G structure and the DDA structures with a different ceramic loss tangent is listed in Table 2. This shows that an average loaded gradient of 100 MV/m can be achieved, with input RF powers of 30.5 MW and 31 MW for an 18-cell DDA structure with different loss tangents of $6 \times 10^{-6}$ and $1 \times 10^{-5}$, respectively, while the RF power is 61.3 MW for the CLIC-G structure. The average powers dissipated in the dielectrics are 198.2 W and 298.8 W with different loss tangents of $6 \times 10^{-6}$ and $1 \times 10^{-5}$, respectively, at these powers and pulse lengths with a repetition rate of 100 Hz.

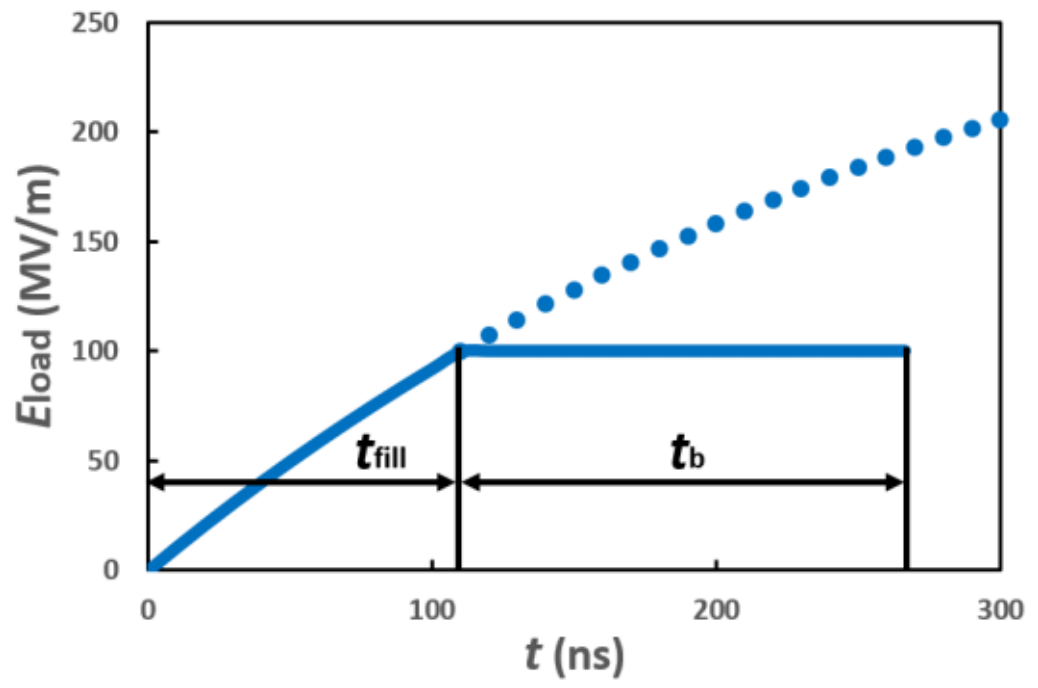

Figure 11. Relationship between the average loaded accelerating gradient $E_{\rm load}$ and time $t$ for a DDA structure.

Table 2. RF parameters of the CLIC-G structure and 18-cell DDA structures with different ceramic loss tangent.

| Geometry | CLIC-G | 18-cell DDA | 18-cell DDA |
|---|---|---|---|
| Dielectric constant $\varepsilon_r$ | | 9.64 | 9.64 |
| Loss tangent tan $\delta$ | | $6\times10^{-6}$ | $1\times10^{-5}$ |
| Loaded gradient [MV/m] | 100 | 100 | 100 |
| Acceleration mode | $TM_{01}$-2π/3 | $TM_{02}$-π | $TM_{02}$-π |
| Shunt impedance [MΩ/m] | 92 | 605 | 527 |
| Peak input power [MW] | 61.3 | 30.5 | 31.0 |
| Filling time $t_{\rm fill}$ [ns] | 67 | 137.6 | 143.9 |
| Beam time $t_{\rm b}$ [ns] | 155.6 | 155.6 | 155.6 |
| RF-to-beam efficiency | 28.5% | 46.6% | 44.8% |

# IV. Transverse Wakefield Analysis

The transverse wakefield is excited when the charged particle beam is not perfectly centred through the accelerating structure, which causes the beam breakup (BBU) instability. The transverse wakefield is found to be proportional to frequency as $w^3$ [53], so it becomes an severe issue for an X-band accelerating structure. It is therefore imperative to study the wakefield behaviour for an X-band DDA structure. In this section, the GDFIDL code [54] is used to compute the time-domain transverse wakefield for a DDA structure as shown in Fig. 12.

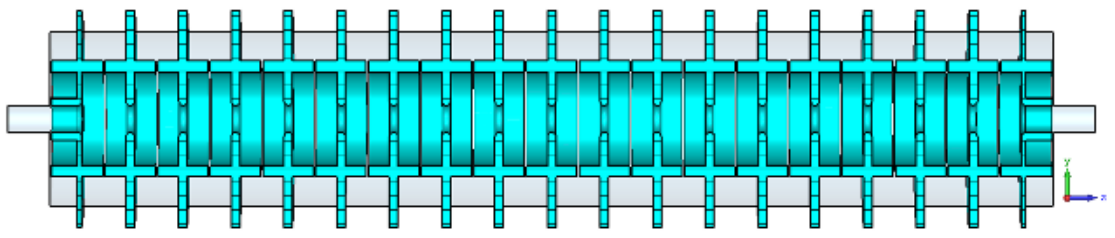

Figure 12. A DDA structure with 18 regular cells and two end cells for GDFIDL simulations.

The wakefield can act both on the bunch itself (called "short-range wakefield") and on successive bunches in the train (called "long-range wakefield"). We will study both cases of the transverse wakefield for a DDA structure. As a comparison, the wakefield is also calculated for a CLIC-G structure with 26 regular cells and two coupler cells. In order to keep the same accelerating length as that of a CLIC-G structure, 18 regular cells and 2 end cells are chosen for a DDA structure, as shown in Fig. 12. A drive bunch with a longitudinal RMS length of $\sigma_z = 1.0$ mm, a bunch charge of $Q$ = 1.0 nC, and an offset of $\Delta d = 0.5$ mm along the $y$ axis, is chosen for the following simulations.

## A. Short-range wakefield

We begin with the analysis of short-range wakefield for both structures. The short-range wakefield is inversely proportional to the 4th power of the iris radius [55]. Thus a small iris aperture imposes an intense short-range wakefield. The radii of the iris aperture are 3.15 mm and 2.35 mm for the first and last structure cells, respectively, for a tapered CLIC-G structure with 28 cells. The DDA structure with 18 regular cells and two end cells has a constant iris aperture radius of 3.15 mm. It was found in Ref. [56] that the very short-range wakefield is not averaged over all irises of the tapered structure; instead it is dominated by the irises with the smallest aperture radius, which is 2.35 mm for a CLIC-G structure. Consequently, the short-range wakefield in the DDA structure can be expected to be lower than that of the CLIC-G structure. This is in good agreement with the simulation results shown in Fig. 13. It should be noted that here the DDA structure has an iris aperture radius of 3.15 mm, which is superior to that of a tapered CLIC-G structure in terms of beam dynamics requirement for practical considerations. In summary, the DDA structure with an iris aperture radius of 3.15 mm has a short-range transverse wakefield lower than that of the CLIC-G structure with tapered iris aperture radii, a great advantage which makes it a strong candidate for future linear accelerators.

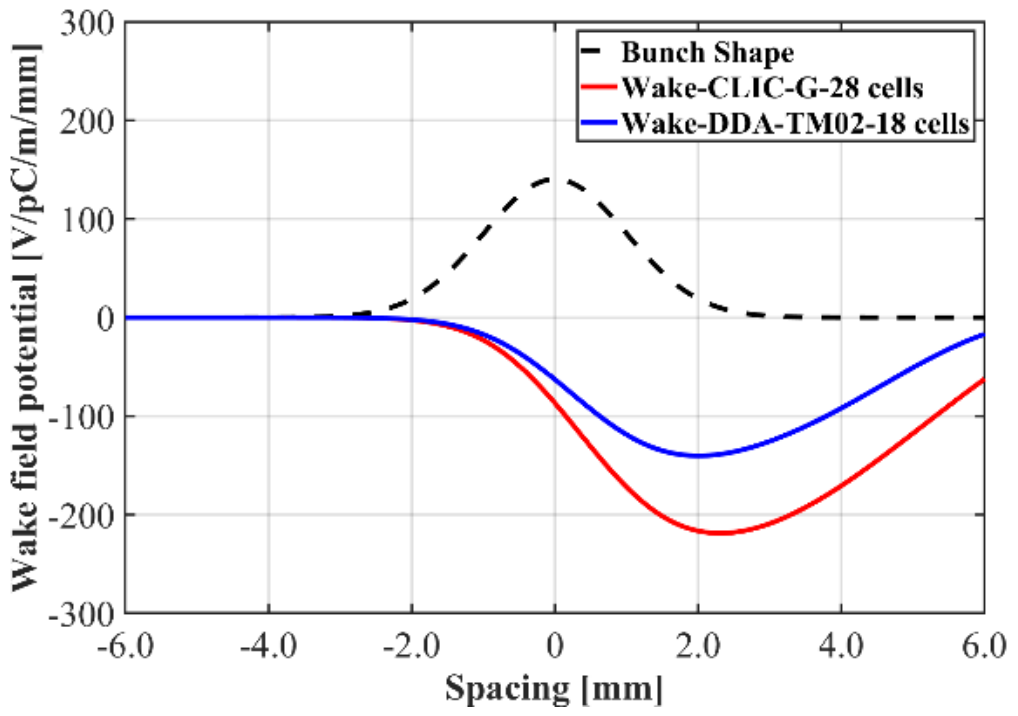

Figure 13. Simulated short-range transverse wakefield for a CLIC-G structure with 28 cells (red curve), and a DDA structure with 18 regular cells and two end cells (blue curve). Black dash line denotes the bunch shape used for simulations.

## B. Long-range wakefield

Damping of the long-range wakefield has been the subject of extensive investigations for normal conducting linear accelerators, such as CLIC. The CLIC main linac accelerating structures usually incorporate cell-to-cell detuning and strong damping, for wakefield suppression. The beam dynamics study for the CLIC main linac indicates that the transverse potential must be suppressed to less than 4.75 V/pC/m/mm [7, 10-11], in order to maintain the beam stability in the main linac. In the GDFIDL simulations, four waveguides with a width $W = 10$ mm, and absorbing boundaries, are added to a DDA structure with 18 regular cells and two end cells, as shown in Fig. 14. The dipole modes are coupled to these waveguides and absorbed by setting the absorbing boundary condition. All the regular cells have the same geometry, so detuning is not included for our simulations at this stage.

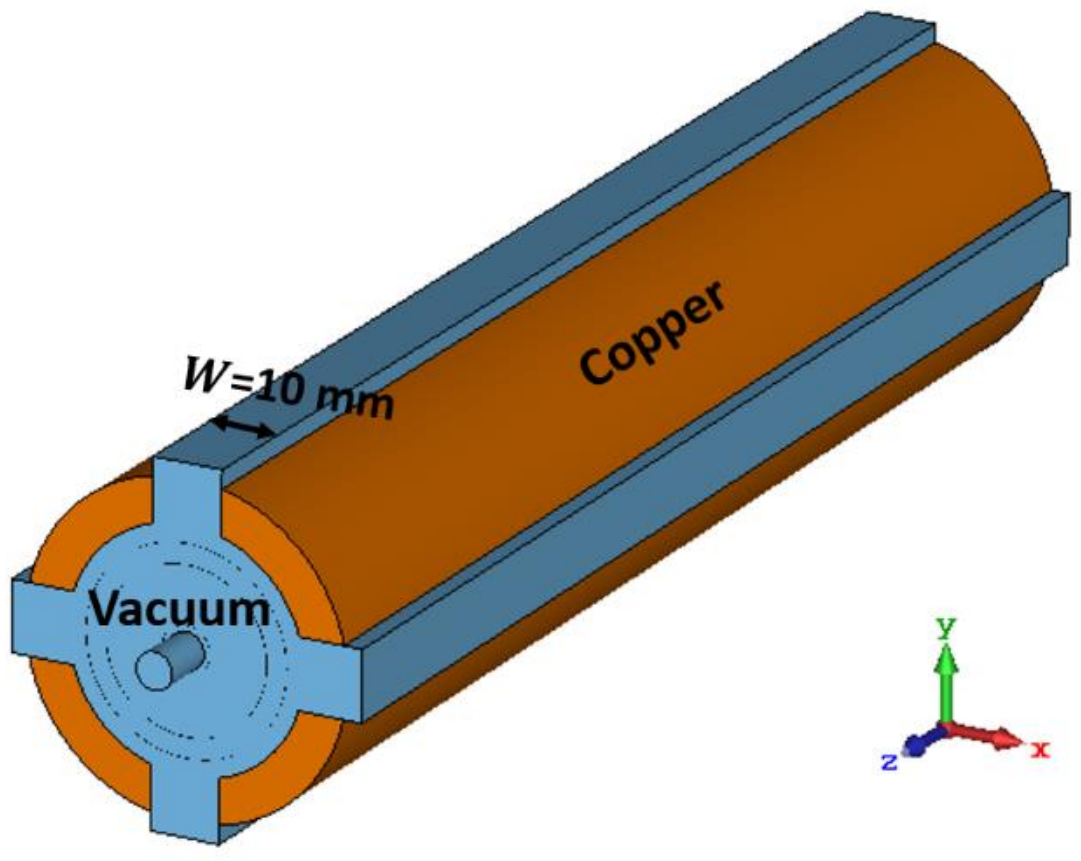

Figure 14. A DDA structure with four absorbing waveguides.

Figure 15 shows the simulated long-range transverse wakefield for a DDA structure and a CLIC-G structure. Schemes for both cell-to-cell detuning and damping are applied for the CLIC-G structure for wakefield suppression, while only damping is only used for the DDA structure. As shown in Fig. 15, the wakefield potential for the CLIC-G structure is damped quickly so that it meets the beam dynamics requirement. However, the wake potential for the DDA structure is damped much more slowly and it remains oscillating without attenuation after a long time as shown in Fig. 15. This is probably because an additional dielectric layer traps the dipole modes inside the structure. This also means that the scheme of adding damping waveguides does not work well. Therefore, other schemes will be essential in a next stage for suppressing the transverse wakefields, to meet the beam dynamics requirement for CLIC.

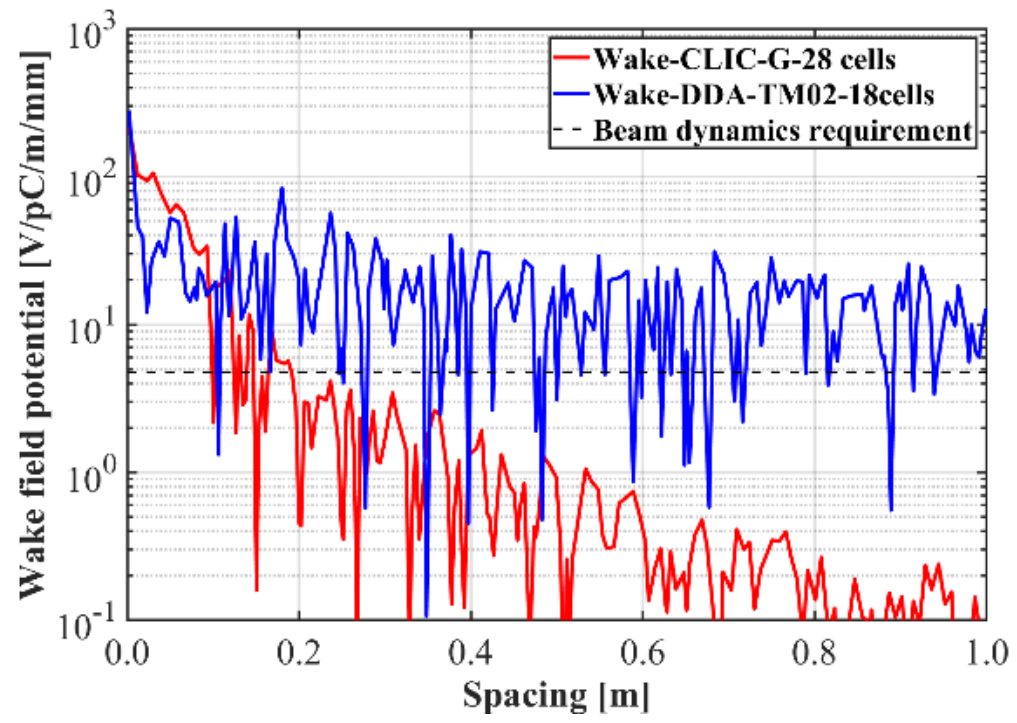

Figure 15. Simulated long-range transverse wakefield for a CLIC-G structure with 28 cells (red curve) and a DDA structure with 18 regular cells and two end cells (blue curve).

# V. Conclusion

This report presents numerical simulations for an efficient X-band DDA structure operating at $TM_{02}$-$\pi$ mode, including optimizations of the regular cell and the end cell, analysis of the accelerating performance of the whole structure, and preliminary wakefield studies. After optimization studies, an extremely high quality factor $Q_0 = 111064$ and a very high shunt impedance $R_{\text{shunt}} = 605$ MΩ/m are obtained for a regular cell. For a single-cell DDA structure with an optimum regular cell electrically shorted just by conducting end plates, two end cells are added to reduce the total RF power loss by about 70%. By analysing the dispersion relation of the accelerating mode, the DDA structure is allowed to have a maximum number of 73 regular cells with a frequency separation of 1.0 MHz to avoid mode overlapping. It is also found that with a larger number of regular cells, the unloaded quality factor and shunt impedance of a DDA structure gradually increase and become saturated to those of a regular cell with periodic boundary conditions. When the number of regular cells for a DDA structure is more than 13, the unloaded quality factor and shunt impedance become constants with $Q_0 = 111064$, and $R_{\text{shunt}} = 605$ MΩ/m, respectively. Consequently, the RF-to-beam power efficiency reaches 46.6%, which is 63.5% higher than previously reported CLIC-G structures with an efficiency of 28.5%. All of these RF parameters are obtained using a ceramic material with an ultralow loss tangent of $6 \times 10^{-6}$. However, the loss tangent may vary and depends strongly on the manufacturing process for the ceramic materials. A lower loss tangent of $10^{-5}$ can be obtained in a manufacturing process which is relatively simple, compared to realizing an ultralow-loss tangent of $6 \times 10^{-6}$ which requires complicated treatment. The corresponding unloaded quality factor and shunt impedance are $Q_0 = 96767$ and $R_{\text{shunt}} = 527$ MΩ/m, respectively, resulting in a RF-to-beam efficiency of 44.8%, for a loss tangent of $10^{-5}$.

In addition, the DDA structure has a short-range transverse wakefield lower than that of CLIC-G structures, which makes it a strong candidate for future linear accelerators. However, adding damping waveguides alone does not work well to suppress the long-range transverse wakefield for a DDA structure. Further investigations are being carried out for damping the wakefields heavily, in order to meet the requirement for high current applications.

# Acknowledgments

The authors would like to thank Dr. Walter Wuensch for the fruitful discussions and useful comments and Dr. Mark Ibison for his careful reading the manuscript.